\documentclass[letterpaper,10pt,oneside,twocolumn]{article}%
\usepackage{geometry}
\usepackage[center]{titlesec}%
\titleformat{\section}[hang]{\normalfont\normalsize\bfseries}{\thesection}{12pt}{\centering}%
\titleformat{\subsection}[display]{\normalfont\normalsize}{\thesubsection}{12pt}{\underline}%
\titleformat{\subsubsection}[runin]{\normalfont\normalsize}{\thesubsubsection}{12pt}{\underline}%
\newcommand{\PaperTitle}[1]{%
   \begin{center}%
      \begin{large}%
         \textbf {#1} \\%
      \end{large}%
   \end{center}%
}%
\newcommand{\AuthorList}[1]{%
   \begin{center}%
      {#1} \\%
   \end{center}%
}%
\newcommand{\AuthorAffiliations}[1]{%
   \begin{center}%
      {#1}%
   \end{center}%
}%
\newcommand{\Keywords}[1]{%
   \begin{center}%
      Keywords: {#1}%
   \end{center}%
}%

\usepackage{graphicx}
\begin{document}%
\twocolumn[%
%
%
\PaperTitle{The effects of Vanadium on the strength of a bcc Fe $\Sigma$3(111)[1$\bar{1}$0] grain boundary}%
%
%
\AuthorList{Sungho Kim$^1$, Seong-Gon Kim$^2$, Mark F. Horstemeyer$^1$,
Hongjoo Rhee$^1$}%
\AuthorAffiliations{$^1$Center for Advanced Vehicular Systems, Mississippi State University, P. O. Box
5405, Mississippi State, MS 39762, USA}%
\AuthorAffiliations{$^2$Department of Physics and Astronomy, Mississippi State
 University, Mississippi State, MS 39762, USA}%
]%
%
%
\section{Abstract} The effects of micro-alloying element, vanadium, on a bcc Fe $\Sigma$3(111)[1$\bar{1}$0]
symmetric tilt grain boundary strength are studied using density
functional theory calculations. The lowest energy configuration of the grain
boundary structure are obtained from the first-principles calculations. The
substitutional and interstitial point defect formation energies of vanadium in
the grain boundary are compared. The substitutional defect is prefered to
interstitial one.  The segregation energies of vanadium onto the grain boundary
and its fractured surfaces are computed. The cohesive energy calculation of the
grain boundary with and without vanadium show that vanadium strengthen the
bcc iron $\Sigma$3(111)[1$\bar{1}$0] grain boundary. 
%
\section{Introduction} %
The macroscopic behavior of steel alloy and their
capabilities for technological applications are vitally influenced by the
properties of micro-alloying elements in their microstructures.
Small amounts of microalloying elements such as vanadium (V), titanium
(Ti), or niobium (Nb) increase strength of steels by grain size control,
precipitation hardening, and/or solid solution
hardening\cite{AHSS:2009,ASM:1990}. V, Nb, and Ti combine preferentially
with carbon and/or nitrogen to form a fine dispersion of precipitated
particles in the steel matrix. Nb may be added in high strength low alloy
(HSLA) sheet to increase the strength predominatedly via grain refinement,
while other microalloys apply the strengthening mechanism of precipitation
hardening to a major extent(Ti) or totally(V)
\cite{Klaus:2002,Klaus:2003}. The addition of small amounts of V increases
the yield strength and the tensile strength of carbon steel. V is one of
the primary contributors to precipitation strengthening in microalloyed
steels. When thermomechanical processing is properly controlled, the
ferrite grain size is refined and there is a corresponding increase in
toughness. V also increases hardness, creep resistance, and impact
resistance due to formation of hard vanadium carbides limiting grain size.
Since V is very effective on aforementioned properties, it is added in
minute amounts. At greater than 0.05\%, however, there may be a tendency
for the steel to become embrittled during thermal stress relief treatments
\cite{Edgar:1939,Joseph:2001}.

The cohesion at these grain boundaries affects the hardness, deformability, and
toughness of the material and it can be enhanced or decreased by segregated
impurities. Therefore, it is essential to understand the interfacial cohesion and
impurity segregation in detail, and a meaningful goal is to find general rules
that describe the relationship between these microscopic features and the
macroscopic properties.

The vanadium(V) is the most common cohesion enhancer that
changes the strength of iron(Fe) metal by the segregation at grain
boundaries. The grain boundary segregation occurs within a few atomic layers at
the grain boundary plane. The grain boundary cohesion enhancement is caused by the
change in the cohesive properties of atoms within a few atomic layers at
the grain boundary plane.

In 1989, Rice and Wang\cite{Rice:1989} developed their theoretical model
for solute segregation into grain boundary and insisted that
the energy required for interfacial separation of grain boundary is the
most important contribution to embrittlement of the grain boundary.
They used the solutes (C, P, S, Sb, Sn) in iron to show 
that the segregation-induced change of separation energy of grain
boundary is roughly consistant with segregation-induced embrittlement.
They estimated the separation energy from  experimental segregation
energies in fractured surface and grain boundary.


A fracture of the grain boundary
creates two separate fractured surfaces under the action of stress. The
cohesive energy of grain boundary $\gamma_{coh}$  (J/m$^2$) is defined as the energy
difference between the energy sum of two surfaces after fracture and the
grain boundary energy before fracture. 

\begin{eqnarray}
	\gamma_{\mathrm{coh}} = 2 \gamma_\mathrm{s} - \gamma_\mathrm{gb}
\label{eq:cohesive}
\end{eqnarray}

where $\gamma_s$  is the surface energy of the two fracture surfaces after fracture,
and $\gamma_\mathrm{gb}$ is the grain boundary energy before fracture.  The cohesive energy in
the presence of segregations of solute atoms can be defined as follows: 

\begin{eqnarray}
 \gamma_{\mathrm{coh}}^\mathrm{seg} & = & ( 2 \gamma_s -
 \frac{NE_s^\mathrm{seg} }{A} ) - ( \gamma_\mathrm{gb} -
 \frac{NE_\mathrm{gb}^\mathrm{seg} }{A} )  \nonumber \\
 & = & \gamma_\mathrm{coh} - ( E_s^\mathrm{seg}  - E_\mathrm{gb}^\mathrm{seg}  ) \frac{N}{A} 
\end{eqnarray}

where $E_s^\mathrm{seg}$ and $E_s^\mathrm{seg}$ are the segregation energy on the surface and
grain boundary. The N is the number of atoms in the unitcell. The A is the
area of surface or grain boundary.

The grain boundary segregation and trapping of vanadium atoms at the fracture
surface affect the cohesive energy of the grain
boundary\cite{Buban:2006,Geng:1999,Yamaguchi:2003,Yamaguchi:2007,GangLu:2005,GHLu:2006,Janisch:2003,Lozovoi:2006,YZhang:2007,Freeman:2000,Olson:2000,Freeman:2001,Schweinfest:2004,Geng:1999,Freeman:1997,Freeman:1996}. If the
fracture surface segregation energy is larger than the grain boundary
segregation energy for a solute element, this element can reduce the cohesive
energy of the grain boundary; it indicates that this element is an embrittling
element. If the grain boundary segregation energy is larger than the
fracture surface segregation energy for a solute element, on the other hand,
this element is a strengthening element. We assume that the total amount of
segregated solute atoms does not change during fracture. This assumption is
valid for elements like vanadium.
\begin{figure}
  \centering
  \resizebox{0.7\hsize}{!}{\includegraphics{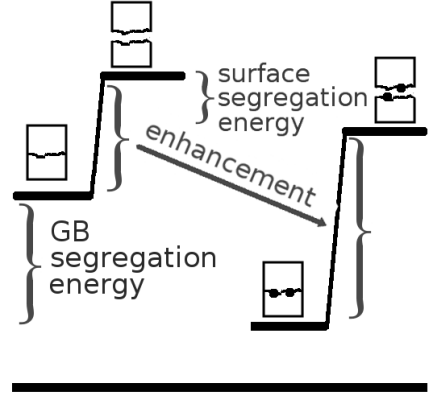}}
  \linespread{1.2}
	\caption{\label{fig:SegEnDia} The schematic of the cohesion
	enhancement effect on the grain boundary by solute atom segregation at the
	grain boundary.}
\end{figure}
Fig.~\ref{fig:SegEnDia} shows the schematic of the
cohesion enhancement effect on a grain boundary by solute
atom segregation at the grain boundary. The energy difference before and after
the grain boundary fracture is affected by solute atom segregation at the grain
boundary. If the segration lower the fracture energy difference, the solute
atom have embrittlement effect on the grain boundary, otherwise cohesion
enhancement effect.

\section{Computational Methods} %
All computation in this paper employed the electronic structure calculations
based on the first principles density-functional theory \cite{Kresse:1996,
  Hohenberg:1964, Kohn:1965}.
The total-energy calculations and geometry optimizations were
performed within density-functional theory (DFT)
\cite{Kresse:1996, Kohn:1965} using the
projector-augmented-wave method.\cite{Blochl:1994,Kresse:1999} All
calculations were spin polarized and the Voskown analysis is used for
the magnetic moment calculations.
The wave function of electrons are expanded in terms of plane-wave basis set
and all plane waves that have kinetic energy less than 250~eV are included
in expanding the wave functions.  For the treatment of electron exchange
and correlation, we use the generalized gradient approximation (GGA) using
Perdew-Burke-Ernzerhof scheme.\cite{Perdew:1996}  For the determination of
the self-consistent electron density 4$\times$1$\times$3 Monkhorst-Pack
k-point set has been used. The structure optimizations were performed
until the energy difference between successive steps becomes less than
$10^{-3}$~eV.  

Fig.~\ref{fig:GBground} shows simulation box and atom configuration. Two
layers are in the periodic cell in $[\bar{1}12]$ direction. The grey
balls represent iron atoms.  Four layers are periodic in
$[1\bar{1}0]$ direction. The Fe bcc $\Sigma 3(111)[1\bar{1}0]$ grain
boundary is indicated with a black arrow.  The total number of
atoms in the simulation box is 88. The periodic boundary conditions are
used in all three direction. Spin is not considered.
\section{Vanadium interaction with the grain boundary and the fractured
surface} %
\begin{figure}
  \centering
  \resizebox{0.7\hsize}{!}{\includegraphics{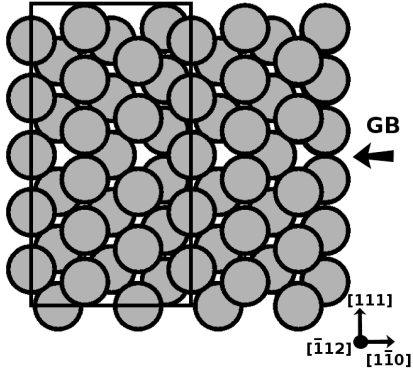}}
  \linespread{1.2}
	\caption{\label{fig:GBground} The optimized grain boundary
	structure of Fe BCC $\Sigma$3(111)[1$\bar{1}$0], and simulation box and
	configuration. There are no vacuum and periodic conditions apply in three
	directions. Two layers are periodic
	in $[\bar{1}12]$ direction. The grain boundary is indicated by a black arrow.
	The atoms in first neighbor layers near the grain boundary are relaxed from
	regular bcc site most significantly. }
\end{figure}

We obtained the optimized grain boundary structure of Fe bcc
$\Sigma$3(111)[1$\bar{1}$0] shown in Fig.~\ref{fig:GBground}.
The atoms in first neighbor layers near the grain
boundary are relaxed from regular bcc site most significantly. The other atoms doesn't
change their positions much in the optimization process.
Our calculated grain boundary formation energy per unit area is 0.113
eV/\AA$^2$ which is the energy required the grain boundary to form from
the Fe bcc bulk.

The grain boundaries usually have many hollow sites bigger than normal
between atoms. The hollow sites are good candidate for micro-alloying
element segregation. In our unitcell there are two hollow sites.
We calculated the vanadium interstitial formation energy in one hollow
site out of two. A vanadium atom is placed at a few different places in a
hollow site to find the best vanadium site to lower the system energy.
\begin{figure}
  \centering
  \resizebox{0.7\hsize}{!}{\includegraphics{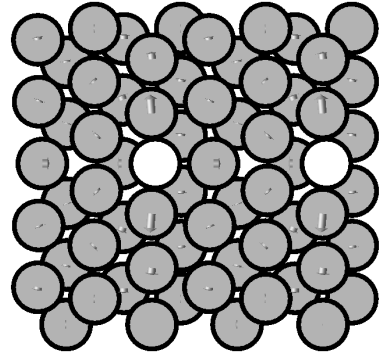}}
  \linespread{1.2}
	\caption{\label{fig:GB_V_grd} The optimized structure of
	interstitially segregated vanadium atom on Fe bcc
	$\Sigma$3(111)[1$\bar{1}$0] grain boundary. The grey balls represent
	iron atoms and white vanadium.  }
\end{figure}
Fig.~\ref{fig:GB_V_grd} show the best interstitially vanadium-segregated
grain boundary structure. The vanadium atom is in the plane of front layer
and shifted a little bit in $[1\bar{1}0]$ direction. The vanadium push away
a little bit the upper and lower neigher iron atoms because the two iron
atoms are too close to the vanadium atom.

The vanadium interstitial formation energy on the grain boundary is
-4.20~eV from a isolated vanadium atom and 1.13~eV from a vanadium atom in
bulk.
The isolated vanadium atom lower the system energy a lot by forming a
interstitial in the grain boundary and forming many bonding with its
neighbor iron atoms.

The micro-alloying element can also be segregated on the grain boundary as a
substitutional form. There are four candidate iron site around grain
boundary for vanadium substitution. The best substitution site is the iron
atom exactly on the grain boundary to lower system energy.
\begin{figure}
  \centering
  \resizebox{0.7\hsize}{!}{\includegraphics{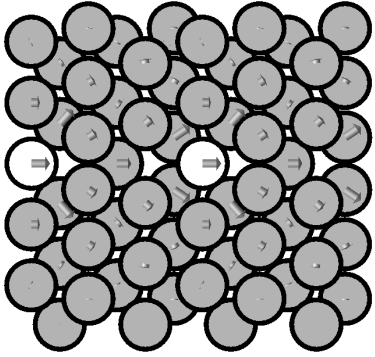}}
  \linespread{1.2}
  \caption{\label{fig:GB_V_sub_g} The optimized structure of
	substitutionally segregated vanadium atom on Fe bcc
	$\Sigma$3(111)[1$\bar{1}$0] grain boundary. The grey balls represent
	iron atoms and white vanadium. }
\end{figure}
Fig.~\ref{fig:GB_V_sub_g} show the substitutionally vanadium-segregated
grain boundary structure of Fe bcc $\Sigma$3(111)[1$\bar{1}$0].  The
vanadium atoms are white to distinguish from grey iron atoms.
The optimized vanadium substituted structure is very similar to the grain
boundary structure without vanadium. 

The vanadium substitutional formation energy on the grain boundary is
-6.08~eV from a isolated vanadium atom and -0.75~eV from a vanadium atom in
bulk. The negative sign means that the vanadium substitution for iron
atom on grain boundary is an exothermic process. Compared to
interstitutional formation energy, the substitutional formation energy is
lower by 1.8~eV which means that substitution defects occur far more often
than interstitial defects in real world. Therefore we only consider the
substitutional segregation at the grain boundary.

The vanadium segregation energy into the grain boundary is the difference
of the grain boundary formation energy from vanadium defect formation
energy in bulk. In order to calculate the vanadium segregation energy into
the grain boundary we calculated two vanadium point defect formation
energies in bulk. One is interstitial point defect formation energies
which are -1.36~eV for tetrahedral interstitial from isolated vanadium
atom and 3.97 from bulk vanadium atom while octahedral interstitial defect
energies are higher by 0.43~eV. The other is substitutional point defect
formation energy which is -6.06~eV from isolated atom and -0.73~eV from bulk
atom. The substitutional defect formation energies in bulk are again lower
than interstitial one by 4.70~eV. From the calculatated energies we can
conclude that vanadium atoms exist mostly as substitutional defects in
bulk and segregate into substitutional defects in grain boundary.

\section{The effect of vanadium on the grain boundary} %
The vanadium segregation energy from bulk to grain boundary is the
substitutional defect formation energy in grain boundary substracted by
the substitutional defect formation energy in bulk. We calculated the
segregation energy which is -0.03~eV. The negative sign mean that the
segregation is an exothermic process.

The surface formation energy of (111) is calculated and  0.167~eV/\AA$^2$
per unit area. The vanadium segregation energy from bulk into surface is
calculated to be -9.57~eV. The negative represent that segregation is an
exothermic process.

The grain boundary cohesion energy(GBCE) defined in Eq.~\ref{eq:cohesive}
without vanadium is calculated as 1.43~eV.
The GBCE with segregated vanadium is 1.54~eV. The vanadium segregation
increases GBCE by 0.11~eV or 0.066~J/m$^2$.


According to our DFT calculation results we can conclude that the
segregated vanadium atom in grain boundary strengthen the grain boundary
against the brittle grain boundary fracture.

%

\section{Summary and Conclusions} %
In summary, we studied the effects of Vanadium on a bcc iron $\Sigma$3(111)[1$\bar{1}$0]
grain boundary strength. We calculated the optimized grain boundary
structure and  the best vanadium segregation site on the grain boundary.
We compared the interstitial defect formation energy to the substitutional
one in bulk and the grain boundary. The substitutional formation energy is
lower than interstitial segregation energy.
Our results indicate that the substitutional segregation is more desirable.
We also calculated vanadium segregation energies on the fractured surface
of the grain boundary.
Based on those segregation energies we calculated the cohesive energy
of the grain boundary and vanadium segregation effect on cohesive energies.

In conclusion, vanadium atom mostly exist as substitutional defects ratner
than interstitial defects in both bulk and grain boundary. Our
first-principle calculations show consistantly with experiment that
vanadium is a Fe grain boundary cohesion enhancer and strengthens the
Fe bcc $\Sigma 3(111)[1\bar{1}0]$ grain boundary.
\bibliographystyle{plain}
\bibliography{DFT}
\end{document}